\documentclass[a4paper,20pt]{article}
    \usepackage[T1]{fontenc}
    \usepackage{graphicx}
    \usepackage{epsfig}
    \usepackage{amsmath}
    \usepackage{amsfonts}
    \renewcommand{\abstract}{}
\begin{document}
\makeatletter
\renewcommand{\@oddhead}{\textit{YSC'14 Proceedings of Contributed Papers} \hfil \textit{A. Golovin et al.}}
\renewcommand{\@evenfoot}{\hfil \thepage \hfil}
\renewcommand{\@oddfoot}{\hfil \thepage \hfil}
\fontsize{11}{11} \selectfont

\title{Monitoring of FR Cnc Flaring Activity}
\author{\textsl{A.Golovin$^{1, 2}$, M.Andreev$^{3, 4}$, E.Pavlenko$^5$, Yu.Kuznyetsova$^2$, V.Krushevska$^2$, A.Sergeev$^{3, 4}$}}
\date{}
\maketitle
\begin{center} {\small $^{1}$Kyiv National Taras Shevchenko
University, Kyiv, Ukraine\\
$^{2}$Main Astronomical Observatory of National Academy of Science
of Ukraine, Kyiv, Ukraine\\
$^{3}$International Center for Astronomical, Medical and Ecological
Research of NASU\\
$^{4}$ Terskol Branch of Institute of Astronomy RAS, Russia\\
$^{5}$Crimean Astrophysical Observatory, Crimea, Nauchnyj, Ukraine\\
astronom\_2003@mail.ru, astron@mao.kiev.ua}
\end{center}

\begin{abstract}
Being excited by the detection of the first ever-observed optical
flare in FR Cnc, we decided to continue photometrical monitoring of
this object. The observations were carried out at Crimean
Astrophysical Observatory (Crimea, Ukraine; CrAO - hereafter) and at
the Terskol Observatory (Russia, Northern Caucasus). The obtained
lightcurves are presented and discussed. No distinguishable flares
were detected that could imply that flares on FR Cnc are very rare
event.
\end{abstract}

\section*{Introduction}
\indent \indent FR Cnc is a BY Dra type star, showing its light
variations (0\hbox{$.\!\!^{\rm m}$}17), caused, as assumed, by the
presence of star-spots and axial rotation \cite{3kazarovets,
4pandeyetal, 5pandey}.

Very recently Golovin, A., Pavlenko, E., Kuznyetsova, Yu. and
Krushevska, V. \cite{2golovin} detected the first ever-observed
large optical flare (1\hbox{$.\!\!^{\rm m}$}02 in B-band,
0\hbox{$.\!\!^{\rm m}$}49 - V-band, 0\hbox{$.\!\!^{\rm m}$}21 -
R-band, 0\hbox{$.\!\!^{\rm m}$}14 - I-band) at CrAO.

FR Cnc has unusually short (for such class of objects) rotational
period of 0.$^d$8267. As it was shown in \cite{1dorren}, short
rotational period allow to expect large flare activity in the star.

In the aim to detect other flares in this object, we decided to
continue photometrical monitoring of FR Cnc.

\section*{Observations}
\indent \indent Observations were carried out by the authors at CrAO
during 4 nights on November, 2006 with the help of 38-cm Cassegrain
telescope and SBIG ST-9 CCD camera, as well as in Terskol
Observatory during 7 nights in March, 2007 using 0.29-m telescope
and Apogee-47 Alta CCD camera (see Table 1 for log of observations).
All observations were done in B-band as far as the flare amplitude
is increasing with decreasing of the wavelength. The duration of
each observational run varies from 2 to 7 hours. Calibration process
of the obtained frames, comparison and check stars are remain the
same as described in \cite{2golovin}.

The obtained lightcurve during the course of FR Cnc monitoring in
March, 2007, folded with FR Cnc rotational period of 0.8267 days, is
shown in Fig. 1.

As it is clearly seen, one-humped 0\hbox{$.\!\!^{\rm m}$}17
variations with the rotational period are clearly distinguishable,
while no flares were detected that could imply that flares is a rare
event for FR Cnc and makes this object even more interesting for
follow-up observations. Taking into account large amplitude of the
flares in short wavelengths, B-band would be recommended,
nevertheless, multicolour CCD photometry is highly valuable as well.

\section*{Acknowledgements}
\indent \indent This work was partly supported by Science \&
Technology Center in Ukraine, grant \#4134.

\vskip 20mm

 {\small \centerline{Table 1. Log of Observations}\nopagebreak
\begin{center}
\begin{tabular}{cccc}
\hline Beginning of the Run (HJD)  &  End of the Run (HJD)  &  Observatory  &  Remarks \\
\hline 2454062.5102 & 2454062.6133 & CrAO & \\
2454065.4893 & 2454065.5540 & CrAO & \\
2454067.4628 & 2454067.5623 & CrAO & \\
2454069.5012 & 2454069.5440 & CrAO & \\
2454171.2453 & 2454171.4361 & Terskol & \\
2454174.3501 & 2454174.4019 & Terskol & \\
2454179.4266 & 2454179.4571 & Terskol & Data excluded due to big scatter \\
2454180.1863 & 2454180.4691 & Terskol & \\
2454181.19303 & 2454181.3536 & Terskol & Data excluded due to big scatter \\
2454182.2395 & 2454182.4972 & Terskol & \\
2454188.4424 & 2454188.4721 & Terskol & \\
  \hline
\end{tabular}
\end{center}
} 

\begin{figure}[!ht]\centering
\begin{minipage}[t]{.75\linewidth}
\centering \epsfig{figure=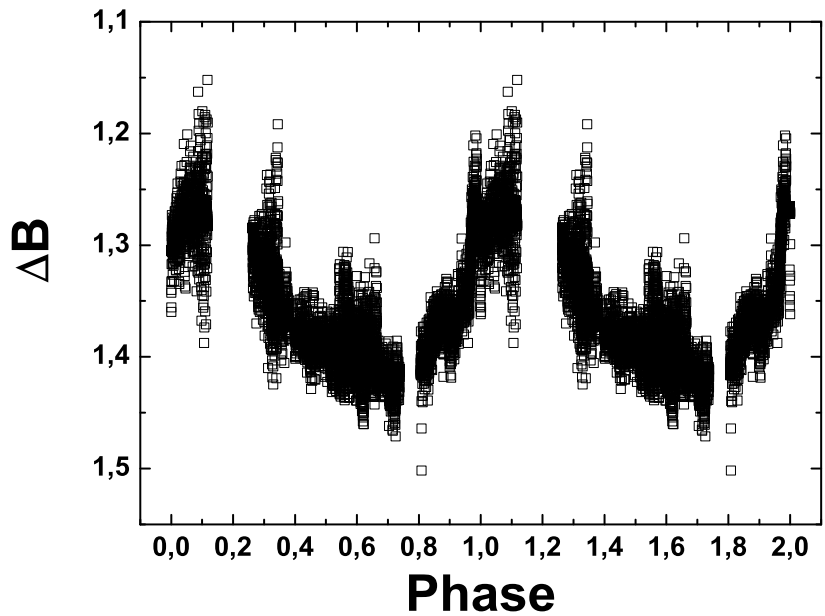,width=\linewidth}
\centering \caption{Phase Lightcurve of FR Cnc.}
\end{minipage}
\end{figure}


\begin{thebibliography}{10}
{\small
\bibitem{1dorren}Dorren J.D., Guinan E.F., Dewarf L.E.  ASPCS, V. 64, p. 399
(1994)
\bibitem{2golovin}Golovin A., Pavlenko E., Kuznyetsova Yu., Krushevska V. IBVS, No.
5748 (2007)
\bibitem{3kazarovets}Kazarovets A.V., Samus N.N., Durlevich O.V., et al. IBVS, No.
4659 (1999)
\bibitem{4pandeyetal}Pandey J.C., Singh, K.P., Sagar R., Drake S.A. IBVS, No.
5351 (2002)
\bibitem{5pandey}Pandey J.C. Bull. of the Astron. Soc. of India, V. 31, p. 329
(2003) }
\end{thebibliography}
\end{document}